%% file: main.tex
\documentclass{article}
\input{formatting.tex}

\title{Resonance Statistics -Informed Fitting Applied to Automated Cross Section Evaluation}

\author{
William Fritsch \and
Noah Walton \and
Justin Loring \and
Jacob Forbes \and
Oleksii Zivenko \and
Aaron Clark \and
Elan Park-Bernstein \and
Vladimir Sobes
}

\date{January 2025}

\begin{document}

\maketitle

\begin{abstract}
This work investigates the use of resonance statistics for resonance evaluation to inform spin group assignment and an alternative fitting objective function beyond the commonly used chi-squared statistic. Resonance statistics -informed methods are applied to the automated resonance fitting framework, developed by N. Walton \textit{et al.} In this automated framework, the utility of resonance statistics is largely unexplored.
The new resonance statistics -informed spin group shuffling algorithm reduces spin group frequency bias seen in the base fitting algorithm. Although resonance statistics -informed optimization produces negligible changes in pointwise cross section agreement, it significantly improves consistency with Wigner level-spacing statistics and stabilizes the fitted resonance density in the presence of model imperfections.
\end{abstract}


\section{Introduction}

Reliable nuclear data form the foundation of nearly all nuclear science and engineering applications. Nuclear data underpins a wide variety of application spaces, including fission and fusion power, nonproliferation, stockpile stewardship, medical isotope production, and astrophysics \cite{NudatNeeds_Bernstein}.
Despite decades of progress, significant pressure remain as progressive nuclear technologies, like Generation IV reactors, push the limits of existing nuclear data. For example, TerraPower traced substantial uncertainty in Molten Chloride Fast Reactor (MCFR) simulations to deficiencies in the $^{35}$Cl(n,p) reaction data, motivating a new measurement campaign at the University of California, Berkeley and a more reliable re-evaluation of the $^{35}$Cl(n,p) cross section \cite{Cl-n-p-exp}.

To address pressures on reliable nuclear data, recent advances in computational capabilities and machine learning have unlocked potential for fast, reliable, and reproducible nuclear data evaluations, but such work is in its infancy \cite{Noah-thesis,Chinese-Auto-Eval}. Automation allows systematic performance testing of various fitting procedures, which will be used to test the hypothesis of this work.

This work concerns \textit{resonance evaluation}, where compiled experimental data are fit using physics-based models such as \textit{R-matrix theory} \cite{R-Matrix}. Resonances -- local amplifications in cross section near specific incident neutron energies -- are identified from structures observed in experimental measurements. While resonance evaluation is primarily guided by agreement with experimental data, resonances also exhibit collective statistical trends known as \textit{resonance statistics} that may be used to enhance performance.

This work explores the following hypothesis: \textit{resonance statistics -informed spin group assignment and fitting will enhance performance, both in reproducing resonance statistics distributions and improving cross section fit}. To test the hypothesis, this work explores resonance statistics in two aspects of the fitting algorithm:
\begin{enumerate}
    \item A Monte Carlo spin group assignment informed by resonance statistics
    \item An objective function informed by both experimental data and resonance statistics
\end{enumerate}

\subsection{Automated Resonance Fitting}
This work builds upon the automated resonance fitting tool of N. Walton, O. Zivenko, and others \cite{Noah-thesis,ND2022_Walton,Noah-2025,Oleksii-metric,ATARI}. This algorithm automates resonance identification using machine learning methods to reduce workload on evaluators. Resonance identifications can be accepted or rejected on the evaluator's discretion.

Several well-established resonance analysis tools exist for tuning resonance evaluations, including SAMMY \cite{sammy} and REFIT \cite{REFIT}, but their capabilities are limited. Given a set of resonance parameters, these tools can:
\begin{enumerate}
\item Evaluate and plot pointwise cross sections
\item Compute theoretical fits and compare them to experimental observables such as transmission and capture yields
\item Adjust resonance parameters to improve agreement with experimental data
\end{enumerate}
These tools require significant input from the evaluator to provide initial resonance parameters and adjust spin group assignments. As a result, resonance evaluation remains an inherently artisanal process, requiring expert judgment to identify resonance energies based on experimental data and prior evaluations. Resonance identification can be exceedingly time consuming and subjective, especially for heavy isotopes that may contain thousands of resonances. While the overarching methodology for resonance evaluation is broadly agreed upon, evaluators often differ in procedural details, leading to variability in evaluations. Reproducibility has therefore become a major concern within the nuclear data community \cite{WPEC-SG49}. The automation resonance fitting algorithm addresses such problems improving reproducibility and reducing workload.

Previous work on the automated resonance fitting tool focused extensively on improving goodness of fit while spin group assignment and resonance statistics distributions remained secondary considerations. In the work of N. Walton \textit{et al.} \cite{Noah-2025}, spin groups were assigned in iteration with a machine-learning spin group reassignment software called BRR \cite{ML_spingroups}; this process was rudimentary and had not been embedded directly within the resonance fitting framework. This work will embed a Monte Carlo spin group assignment algorithm directly into the fitting routine.

\subsection{Outline}
This work is separated into 5 additional sections. Section \ref{sec:background} focuses on background regarding R-matrix theory and resonance statistics. Section \ref{sec:fitting_methods} introduces the spin group selection algorithm and resonance statistics -informed fitting. Section \ref{sec:syndat} introduces metrics for measuring fitting performance. Section \ref{sec:results} applies the resonance statistics methods to an analysis of $^{181}$Ta, measuring performance based on the established metrics. Finally, Section \ref{sec:conclusion} reviews the hypothesis, establishing the best methods based on the defined performance metrics.

\section{Background}\label{sec:background}
\subsection{Resonance Analysis and Evaluation}

Neutron cross sections are a crucial component of nuclear data evaluation. Neutron cross sections describe the likelihood of a reaction occurring from a neutron incident on some target nucleus.
In the energy domain relevant to this work -- the \textit{resolved resonance region} (RRR) -- cross sections are described using mathematical models derived from \textit{R-matrix theory} \cite{R-Matrix}. R-matrix theory models nuclear reactions with discrete, quantized excitation states. When the incident neutron energy is close to one of these states, the probability of interaction is amplified, producing a feature known as a \textit{resonance}.
A substantial portion of a resonance evaluator's effort is devoted to identifying resonance features in experimental data, justifying their physical existence, and determining appropriate resonance parameters. Depending on the isotope and energy range, evaluators may need to analyze tens to thousands of resonances.

In R-matrix theory, a resonance can be characterized by three main components: 1) resonance energy, 2) partial widths, and 3) spin group assignment. Resonance energy ($E_\lambda$) determines the `location' of a resonance in incident neutron energy. Partial widths determine the height and width of resonances under various nuclear reactions. There exists a partial width for each ingoing and outgoing reaction state, called a \textit{channel}. For many isotopes, only two channels per resonance are of particular concern: a neutron width ($\Gamma_n$) and an aggregated capture (gamma) width ($\bar{\Gamma}_\gamma$). Often, many capture channels exist, but they are commonly aggregated into a single channel under the Reich-Moore approximation \cite{ReichMooreApprox}.

Spin group assignment is a collection of quantum numbers describing the spin and angular momentum of the ingoing and outgoing particles. Each resonance has a spin group state. A spin group assignment is defined by a total angular momentum ($J$) and a parity ($\pi$), denoted together as $J^\pi$. Spin group assignment plays a role in the shape of resonances through the spin statistical factor $g_J$. Additionally, spin group assignment determines the interplay of nearby resonances. Resonances with the same spin group will exhibit constructive or destructive interference in the region between resonances, while resonances with different spin groups will simply be additive in cross section space. Spin group assignment also influence the size and frequency of resonances under resonance statistics, motivating the next discussion.

\subsection{Resonance Statistics}
Resonances belonging to the same spin group exhibit systematic statistical behaviors, collectively referred to as resonance statistics \cite{level-density-analysis}. One statistical trend regards the spacing of resonances in energy.
The distribution of energies for resonances of the same spin group resemble the eigenvalues of random matrices drawn from a \textit{Gaussian Orthogonal Ensemble} (GOE).
Such eigenvalues exhibit a sort of pressure between resonances such that resonances are not often found too close nor too far apart. The spacing between neighboring resonances of the same spin group -- commonly referred to as level spacing ($D$) -- follows \textit{Wigner distribution} \cite{wigner_distr_original}. The probability density function for Wigner distribution is given by Equation \eqref{eq:Wigner}, where $\langle D\rangle$ denotes the mean level spacing for the spin group.
\begin{equation}\label{eq:Wigner}
\Pr(D)dD = \frac{\pi}{2\langle D\rangle^2}D\exp\left(-\frac{\pi}{4}\left(\frac{D}{\langle D\rangle}\right)^2\right)dD
\end{equation}
This spectral pressure extends beyond nearest neighbors: short spacings are often followed by large spacings. The consequence is a strong coherence to a uniform resonance density over sufficiently large energy ranges. Evaluators often plot the cumulative number of resonances identified versus energy which follows closely with a linear trend.
Deviation in the cumulative level distribution from expectations -- usually from missing resonances -- is often used as a rough guide for placing the boundary between the RRR and the \textit{unresolved resonance region} (URR).

Another statistical distribution concerns resonance widths. \textit{Observed widths} tend to scale with energy. After correcting observed widths with an energy-dependent penetrability factor $P(E)$, one obtains \textit{reduced widths}, defined by Equation \eqref{eq:penetrability}.
\begin{equation}\label{eq:penetrability}
    \gamma_c^2 = \frac{\Gamma_c}{2P(E)}
\end{equation}
Reduced widths for a given reaction channel follow the Porter-Thomas (PT) distribution \cite{PT_original}, which is a rescaled chi-squared distribution, as shown in Equation \eqref{eq:Porter-Thomas}. A reduced width for a channel $c$ is denoted $\gamma_c^2$. The mean reduced width $\langle\gamma_c^2\rangle$ is assumed constant in the RRR.
\begin{equation}\label{eq:Porter-Thomas}
\Pr(\gamma_c^2)d\gamma_c^2 = \left(\frac{\nu}{2\langle\gamma_c^2\rangle}\right)\frac{\left(\frac{\nu\gamma_c^2}{2\langle\gamma_c^2\rangle}\right)^{\nu/2-1}\exp\left(-\frac{\nu\gamma_c^2}{2\langle\gamma_c^2\rangle}\right)}{\Gamma(\nu/2)}d\gamma_c^2
\end{equation}
The channel multiplicity $\nu$ represents the number of channels aggregated into a single effective width. Channel aggregation is commonly employed to reduce the R-matrix equations to a tractable number of parameters; the Reich-Moore approximation is a notable example of this approach, aggregating all capture widths into a single width.
Gamma decay often involve many open channels, each describing unique decay and spin states, resulting in a large channel multiplicity. As the multiplicity increases, the PT distribution approaches a delta function centered at the mean reduced width. Consequently, evaluators often fix capture widths to a constant mean value \cite{238Pu-Eval}. This work applies the same assumption, fixing the capture widths.
Neutrons often have only one or two identifiable channels per spin group. This work will only consider neutron widths representing one channel.

The quantity $\gamma_c$ is commonly referred to as \textit{reduced width amplitude} and is the fundamental quantity appearing in R-matrix equations. The sign of the reduced width amplitude determine constructive or destructive interference between resonances of the same spin group.


\subsubsection{The Utility of Resonance Statistics}
Resonance statistics play a central role in nuclear data evaluation. In the URR, resonance realizations are sampled from resonance statistics to generate probability tables \cite{SESH-URR-generation,FRENDY-URR-generation}.  In the present work, resonance statistics are essential for three primary applications:
\begin{itemize}
    \item \textbf{Synthetic data generation:} sampling from resonance statistics enables the creation of synthetic cross section realizations used to assess the performance of the automated fitting methodology, as discussed in Section \ref{sec:syndat}.
    \item \textbf{Spin group assignment:} more direct measurements of spin group assignments -- gamma-ray emission, polarized neutron data, or neutron scattering angle measurements -- are uncommon and often expensive \cite{ML_spingroups}. Instead, resonance statistics alone is often used to suggest spin group assignment using tools such as BRR \cite{ML_spingroups} or SUGGEL \cite{suggel}. This work leverages resonance statistics for spin group assignment, discussed in Section \ref{sec:spin_shuffle}.
    \item \textbf{Alternative objective function:} resonance statistics provide probabilistic information about plausible spin group assignments and resonance parameters. This work introduces an objective function informed by resonance statistics, discussed in Section \ref{sec:obj_func}. While resonance statistics in spin group assignment is commonplace, using resonance statistics for optimization is largely unexplored in resonance evaluation motivating the study.
\end{itemize}

\section{Fitting Methods using Resonance Statistics}\label{sec:fitting_methods}
The automated resonance evaluation framework is complex and the full details are beyond the scope of a journal article. The fitting algorithm is well described in N. Walton's thesis \cite{Noah-thesis}. However, many changes have been made to the fitting algorithm since to support this work and the general development of the automated evaluation framework. Appendix \ref{appendix} provides a simplified overview of the fitting algorithm with detailed changes since the thesis. This section focuses on changes to the fitting algorithm primarily involving resonance statistics: spin group assignment and an alternative objective function.

A basic understanding of the fitting algorithm is required going forward. The automated resonance fitting tool that internally uses SAMMY for local optimization.
Automated resonance fitting tool can be separated into three phases: initial solve, stepwise elimination, and model selection. The algorithm begins by constructing an intentionally overfit model during the initial solve phase. This overfit model is designed to represent all physically real resonances while also including a large number of fake resonances. In the stepwise elimination phase, statistically unsupported resonances are removed one at a time. In the elimination phase, a resonance is selected for removal if the model has the most optimal objective value after optimization. The process of selecting the best resonance to eliminate is called \textit{variable selection}. Finally, the \textit{model selection} phase determines the number of resonances retained in the final solution.

\subsection{Spin Group Selection}\label{sec:spin_shuffle}
The automated fitting methodology described by N. Walton and O. Zivenko largely neglects spin group assignment \cite{Noah-thesis}. Spin groups are initially assigned early in optimization, but their impact on the cross section is primarily manifested through resonance-resonance interference and minor shape differences, which becomes significant only in later stages of the fitting algorithm.

To address spin group assignment, a Monte Carlo spin group shuffle algorithm is applied after selected elimination steps. This algorithm generates 5 candidate models, with alternative spin group assignments, weighted by their PT and Wigner likelihoods. Each model is independently optimized and the model yielding the optimal objective value is chosen. The choice of 5 candidate models will be supported by the demonstration in Section \ref{sec:convergence}.
The spin group shuffle algorithm is not applied early in the elimination phase because of the number of false resonances overwhelms the Monte Carlo algorithm. Instead, spin group shuffling is reserved for use in the realm of plausible model cardinalities.\footnotemark{} 

\footnotetext{For $^{181}$Ta, there is a very small chance of GOE-generated ladders within an 11.5 eV window having more than 9 resonances. Therefore, in the application of the algorithm to $^{181}$Ta, Monte Carlo spin group shuffling is only used in elimination for models with 9 resonances or less.}

In the future the spin group shuffling algorithm may be used to shuffle the sign for the reduced width amplitudes, considering different interference patterns. However, under the Reich-Moore formalism, the signs for neutron and capture width amplitudes do not affect the calculated cross section. Width sign shuffling may be used when a new reaction channel is opened -- such as a fission channel -- where the width signs result in tangible changes in cross section.

\subsection{Objective Function and Resonance Statistics}\label{sec:obj_func}


In previous work, the automated fitting algorithm used the chi-squared goodness of fit ($\chi^2$) as the objective function. The $\chi^2$ metric captures experimental consistency, but neglects statistical plausibility of the resonance parameters. This work studies an alternative objective function that augments $\chi^2$ with resonance statistics, encoding statistical plausibility directly into the objective function.

Let $D$ and $V$ denote the experimental data vector and data covariance matrix, respectively. Let $P$ represent the resonance parameters and $T(P)$ the corresponding model prediction evaluated on the experimental data grid. The $\chi^2$ statistic is defined as the Mahalanobis distance between the data and the model prediction,
\begin{equation}\label{eq:chi2}
    \chi^2=(D-T(P))^TV^{-1}(D-T(P))
\end{equation}
The $\chi^2$ statistic is well suited for optimization due to its direct connection to the likelihood of observing Gaussian-distributed data $D$ given a model defined by resonance parameters $P$,
\begin{equation}\label{eq:chi2-likelihood}
    \Pr(D|P)\propto\exp\left(-\frac{1}{2}\chi^2\right)
\end{equation}
The quantity of interest for inference and optimization, however, is the posterior probability of the resonance parameters given the data, $\Pr(P|D)$. By Bayes’ theorem,
\begin{equation}\label{eq:Bayes_fit}
    \Pr(P|D)\propto \Pr(P)\cdot \Pr(D|P)
\end{equation}
$\Pr(P)$ is the prior probability distribution on the resonance parameters. If $\Pr(P)$ is chosen to be infinite uniform, maximizing the posterior probability is equivalent to minimizing $\chi^2$. Alternatively, the prior $\Pr(P)$ may be constructed from resonance statistics, allowing known statistical properties of resonance parameters to inform the fit.
In this work, resonance statistics are incorporated by modifying the $\chi^2$ objective function to
\begin{equation}
    \text{obj} = \chi^2-2\log\left(\Pr(P)\right)
\end{equation}
The specific form of these probability distributions must be chosen carefully. Consider neutron widths in the case of a single channel ($\nu=1$). The reduced neutron widths $\gamma_n^2$ follow PT distribution, whereas the reduced width amplitudes $\gamma_n$ follow a normal distribution:
\begin{equation}
    \Pr(\gamma_n)d\gamma_n = \frac{1}{\sqrt{2\pi\langle\gamma_n^2\rangle}}\exp\left(-\frac{\gamma_n^2}{2\langle\gamma_n^2\rangle}\right)d\gamma_n
\end{equation}
Although these two parameterizations are related by a change of variables, they provide different probability densities and therefore different maximum a posteriori (MAP) estimates. This lack of invariance of MAP estimators under reparameterization is a well-known issue in Bayesian statistics \cite{MAP_variance}.
From a practical perspective, the PT distribution on $\gamma_n^2$ exhibits a divergent probability density as $\gamma_n^2 \to 0$. When using this distribution, the objective function strongly biases small resonances toward vanishing widths. In contrast, the distribution on the reduced width amplitude $\gamma_n$ is Gaussian for $\nu = 1$, resulting in a smooth, well-behaved objective function with finite curvature near the origin. For this reason, the prior is constructed using probability densities on $\gamma_n$ rather than on $\gamma_n^2$. This choice retains the statistical context implied by PT distribution while yielding a smooth, well-behaved objective function that avoids runaway behavior during optimization.
For resonance energies, the Wigner distribution (Equation \ref{eq:Wigner}) provides a natural choice of prior probability density.

Model selection involves comparing posterior probabilities for models with different numbers of resonances. A complication arises because models with different resonance counts are defined on parameter spaces of different dimensionality. With 5 resonances, the joint PT distribution domain is 5-dimensional, one for each resonance. To enable meaningful comparison, \textit{virtual resonances} are introduced so that all models are evaluated in a common parameter space. Under this convention, virtual resonances are unconstrained by the experimental data and therefore adopt parameter values that maximize the prior probability: $\gamma_n=\gamma_n^2=0$ and $D=\langle D\rangle\sqrt{2/\pi}$.
Virtual resonances therefore contribute constant factors to the posterior probability, with corresponding PT and Wigner likelihoods of $1/\sqrt{2\pi\langle\gamma_n^2\rangle}$ and $\sqrt{\pi/2e}/\langle D\rangle$, respectively.
This construction provides a consistent reference point for comparing models of differing cardinality, but prefers more parsimonious models by design. Perhaps, future work may yield an unbiased convention for a consistent parameter space.

The introduction of resonance statistics is expected to have a modest effect when experimental data are dense, since the measurement data dominates the objective function. Take J. Brown $^{181}$Ta data as an example \cite{Ta181_measurements_Brown}. An energy window from 200 eV to 220 eV has an average of about 5 resonances and 621 data points. Since there are over 100 data points per resonance, $\chi^2$ carries most of the statistical weight in the new objective function. However, at 2000 eV, there are only around 25 data points per resonance. At high energies, experimental data becomes more sparse, and resonance statistics has a larger influence the fitting routine.

This work tests modifications to the automated fitting framework using resonance statistics -informed objective functions across three application areas: variable selection, model selection, and gradient-descent optimization. Performance is assessed through four progressively augmented configurations, in which the traditional $\chi^2$ objective function is incrementally replaced by the proposed resonance statistics -informed objective:
\begin{enumerate}
    \item No resonance statistics -informed fitting
    \item Resonance statistics applied to variable selection only
    \item Resonance statistics applied to both variable and model selection
    \item Resonance statistics applied to variable selection, model selection, and gradient-descent optimization
\end{enumerate}
This staged approach isolates the impact of resonance statistics within each component of the automated fitting procedure.

Resonance statistics -informed gradient-descent requires optimization outside of SAMMY's internal generalized least linear squares Bayesian solver. The external solver -- discussed in section 4.2.4 of N. Walton's thesis \cite{Noah-thesis} -- incurs an approximate 50\% runtime penalty due to text file -based derivative handling. Such a runtime could be reduced significantly using the SAMMY C++ or Python API that is under development at the time of this work \cite{SAMMY_ext}. Future work may integrate the Python API into the automated fitting framework if external solvers prove useful.

\section{Performance Measures and Synthetic Data}\label{sec:syndat}
Quantitative performance measures are used to evaluate the effectiveness of the proposed fitting methodologies. It is important to distinguish a \textit{performance metric} from an \textit{objective function}. Performance metrics quantify the performance of a certain aspect of a model. An objective function is an optimization tool, serving as a surrogate for the performance metrics.
Unlike objective functions, performance metrics can be defined relative to an underlying true solution. For real data, the underlying true cross section is unknown, but for synthetically created data, the true cross section is known exactly. A synthetic data framework was developed by N. Walton et al. \cite{Noah-Syndat}, and corresponding performance metrics were formalized by O. Zivenko et al. \cite{Oleksii-metric}. This work adopts these established metrics to quantify the performance of spin group shuffling and resonance statistics -informed objective functions. Some new metrics will be defined to quantify performance in other aspects of the fitting algorithm.
There are two aspects of interest that the performance metrics are designed to capture:
\begin{enumerate}
    \item \textbf{Fitting similarity:} agreement between the fitted cross section and the true cross section.
    \item \textbf{Resonance statistics:} consistency of evaluated resonance parameters with expected statistical distributions.
\end{enumerate}

\subsection{Fitting Similarity Metrics}

O. Zivenko \textit{et al.} \cite{Oleksii-metric} created the foundation for the \textit{root mean squared error} (RMSE) metric to quantify pointwise agreement between the fitted and true cross sections. For a single reaction channel, the mean squared error (MSE) is defined as
\begin{equation}
\label{eq:MSE_one_react}
\text{MSE}_\text{rxn} = \frac{{ \int_{E_{\text {min}}}^{E_{\text {max}}} (\sigma_{\text {fit}}^{\text {rxn}} - \sigma_{\text {true}}^{\text {rxn}})^2 \, dE}}{\int_{E_{\text {min}}}^{E_{\text {max}}}dE}
\end{equation}
For this work, MSE values for elastic and capture reactions are combined as
\begin{equation}
\label{eq:MSE}
\text{MSE} = \frac{1}{N_\text{rxn}}\sum_\text{rxn}\text{MSE}_\text{rxn}
\end{equation}
and the RMSE is defined as the square root of the resulting MSE and is measured in units of barns.
RMSE is the primary metric for the lower energy regions of the RRR, where accurate reproduction of resonance shapes is essential for neutron transport and shielding applications.

The $\chi^2$ goodness of fit is not an ideal performance metric due to its over-dependence on the data and bias towards overfit models, but it does not require knowledge of the true cross section, making it the choice metric for real data.

\subsection{Resonance Statistics Metrics}
Although individual resonance misassignments often have a modest effect on resonance-resonance interference, systematic biases in spin group assignment can accumulate and distort the resonance background. Assessing the statistical consistency of evaluated resonance parameters is essential when introducing spin shuffling and resonance statistics -informed objective functions.
Three resonance statistics-based measures are considered:
\begin{enumerate}
\item \textbf{Empirical level density:} the number of resonances per unit energy, used to assess systematic underfitting or overfitting.
\item \textbf{Wigner level-spacing statistics:} agreement with expected spacing distributions within a spin group.
\item \textbf{Porter-Thomas width statistics:} agreement between fitted and expected reduced width amplitude distributions.
\end{enumerate}

Goodness-of-fit to the Wigner and PT distributions is quantified using the Kolmogorov-Smirnov (KS) \cite{KS_test}, Cramér-von Mises (CvM) \cite{CvM_test}, and Anderson-Darling (AD) \cite{AD_test} statistics.
These metrics compare the empirical cumulative distribution function (ECDF), $\hat{F}(x)$, to the expected cumulative distribution function, $F(x)$, as defined in Equations \eqref{eq:KS_criterion}, \eqref{eq:CvM_criterion}, and \eqref{eq:AD_criterion}.
\begin{equation}\label{eq:KS_criterion}
    \text{KS} = \max\left(\hat{F}(x)-F(x)\right)
\end{equation}
\begin{equation}\label{eq:CvM_criterion}
    \text{CvM} = n\int_0^\infty\left(\hat{F}(x)-F(x)\right)^2dF(x)
\end{equation}
\begin{equation}\label{eq:AD_criterion}
    \text{AD}^2 = n\int_0^\infty\frac{\left(\hat{F}(x)-F(x)\right)^2}{F(x)\cdot\left(1-F(x)\right)}dF(x)
\end{equation}
The AD statistic places greater emphasis on distribution tails, the KS statistic emphasizes the bulk of the distribution, and the CvM statistic provides an intermediate balance. In all cases, smaller values indicate closer agreement with the expected resonance statistics.

The empirical reduced width amplitude distribution will be normalized relative to the fitted frequency over the expected frequency instead of being normalized to 1, as shown in Figures \ref{fig:PT_ecdf_fake_1.5keV} and \ref{fig:PT_ecdf_real_1.5keV}. In this case, AD is divergent and will be not be used for PT distribution.

The three statistics are used best for relative comparisons between models, but one can also test the plausibility of the fit by comparing to the fit from random samples.
A significance of 1\% will be used to determine the plausibility of an empirical distribution. Approximately 99\% of random samples from distributions have a KS value less than $1.63/\sqrt{N}$, where $N$ is the number of parameters for the distribution \cite{KS_test}. For the same significance and for many samples, AD and CvM have 1\% significance thresholds of $1.97$ and $0.743$, respectively.

\section{Application to Tantalum-181}\label{sec:results}
This work tests 5 models on synthetic $^{181}$Ta data and real $^{181}$Ta data. The first model is the baseline without resonance statistics or spin group selection. The rest of the models use spin group shuffling and follow the 4 cases in Section \ref{sec:obj_func}.
First, the experimental data will be discussed, followed by a convergence study on the spin group shuffling algorithm. Next, the performance will be evaluated on both synthetic and real data.

\subsection{Experimental Data}\label{sec:experimental_data}
This work uses the $^{181}$Ta experimental measurements reported by J. Brown \cite{Ta181_measurements_Brown,BrownThesis}, consisting of three transmission measurements and two capture experiments performed at Rensselaer Polytechnic Institute (RPI). These datasets have been used previously for testing and benchmarking the automated resonance fitting algorithm due to their quality, relevance, and recency \cite{Noah-2025,Oleksii-metric,Jake-thesis}.
Radiative capture cross sections for tantalum are of high interest for nonproliferation applications, particularly in gamma-ray spectroscopy \cite{Zr-nuc-data-needs}. In addition, tantalum’s high melting point and corrosion resistance make it a common material for crucibles containing molten actinides. The radiologically active nature of molten actinides places stringent demands on tantalum nuclear data for criticality safety analyses \cite{Ta181_measurements_Brown}.

The resolved resonance region of the ENDF/B-VIII.1 $^{181}$Ta evaluation \cite{ENDF8p1} was fitted using a larger set of experimental data, including transmission measurements by J. Harvey \cite{Harvey_Ta181}. For a fair $\chi^2$ comparison, the ENDF/B-VIII.1 resonance parameters will be optimized locally to fit only the 5 J. Brown datasets used in this work.

\subsection{Spin Group Shuffling Convergence}\label{sec:convergence}
\begin{figure}
    \centering
    \includegraphics[width=0.9\linewidth]{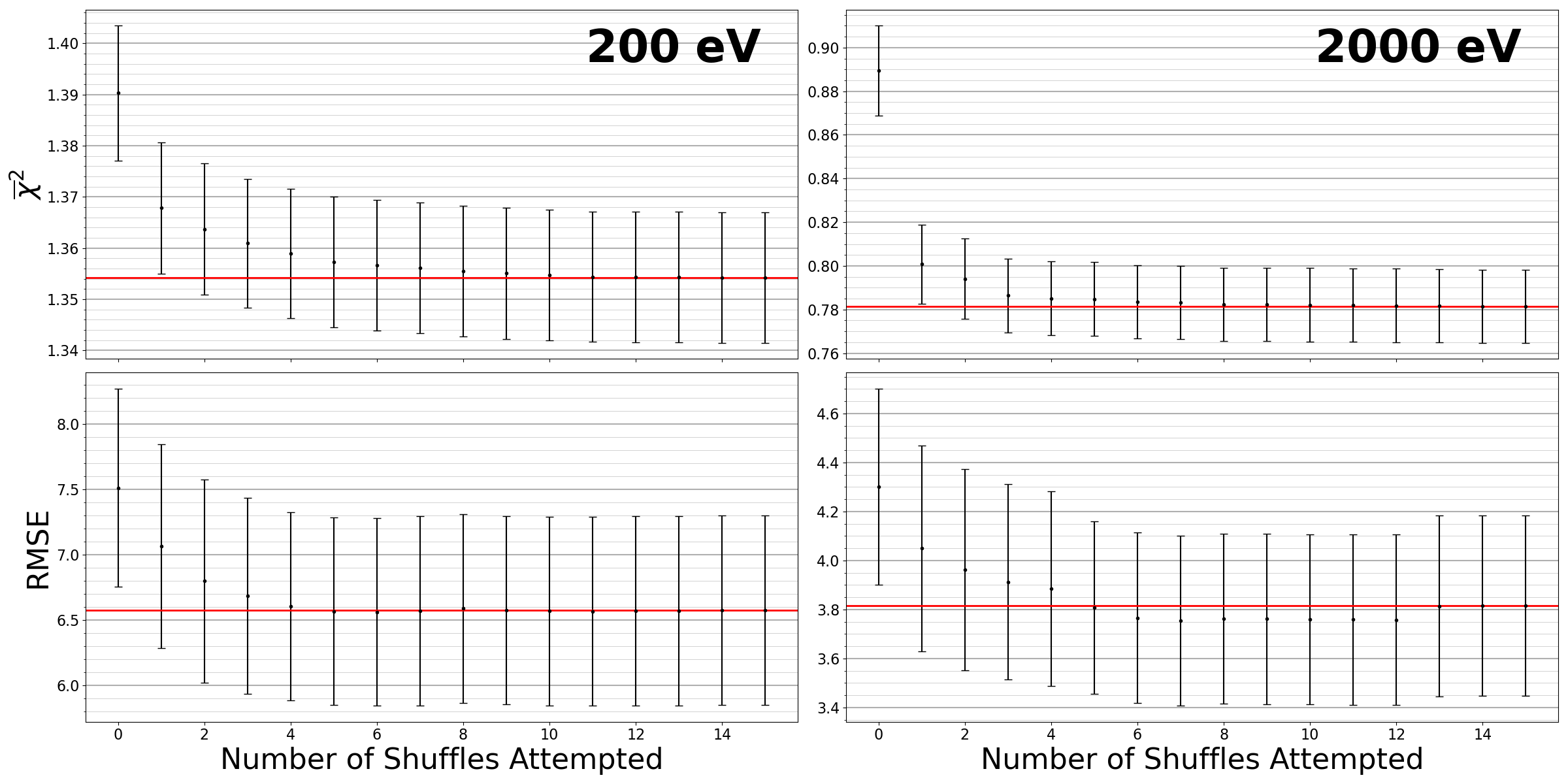}
    \caption{A study on the number of spin group assignment shuffles needed for convergence of an evaluation in the energy range of 200-220 eV (left) and 2000-2020 eV (right). The chi-squared per degrees of freedom and RMSE metric are compared against the number of shuffles. The error bars present the mean and standard deviation of the mean across 100 ladder fits. 5 shuffle attempts were determined sufficient for convergence.}
    \label{fig:shuffle-convergence}
\end{figure}

To determine the number of spin group shuffles required per elimination stage, a convergence study was performed using 100 synthetically generated $^{181}$Ta resonance ladders. In each case, the optimal shuffled ladder was selected based on the minimum $\chi^2$ value. Figure \ref{fig:shuffle-convergence} illustrates the mean and the standard error of the mean for both the $\chi^2$ per data point and the RMSE across two energy ranges: 200–220 eV and 2000–2020 eV. The red lines represent the asymptotic $\chi^2$ per data point and RMSE. In both energy ranges, convergence is reached rapidly, after approximately 5 shuffles. An additional shuffle would generate a marginally better $\chi^2$ for about 10\% of cases. Based on this study, 5 spin group shuffles are used at each elimination stage in the remainder of this work. Although the choice of the number of shuffles may appear low, the spin group shuffling algorithm is reapplied after each elimination, providing many attempts for optimization.

\subsection{Performance on a Synthetic Data Ladder}
Synthetic data were sampled from 150 eV to 3.5 keV, replicating the 5 experimental datasets measured by J. Brown \cite{BrownThesis}. Because the true underlying resonance parameters and cross sections are known, synthetic data allow direct quantification of model performance.

\begin{figure}
    \centering
    \includegraphics[width=0.45\linewidth]{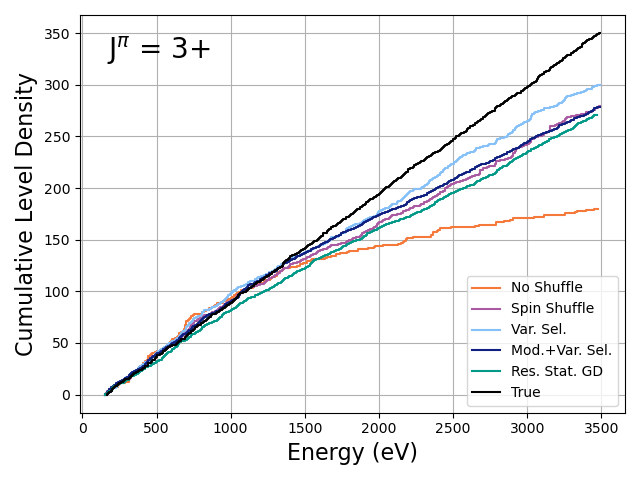}\includegraphics[width=0.45\linewidth]{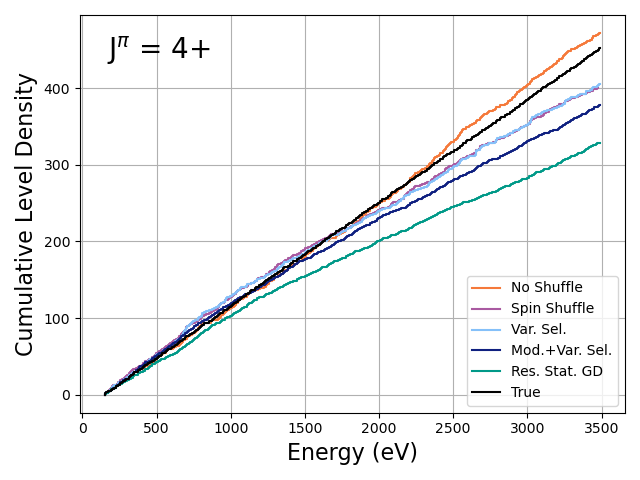}
    \includegraphics[width=0.9\linewidth]{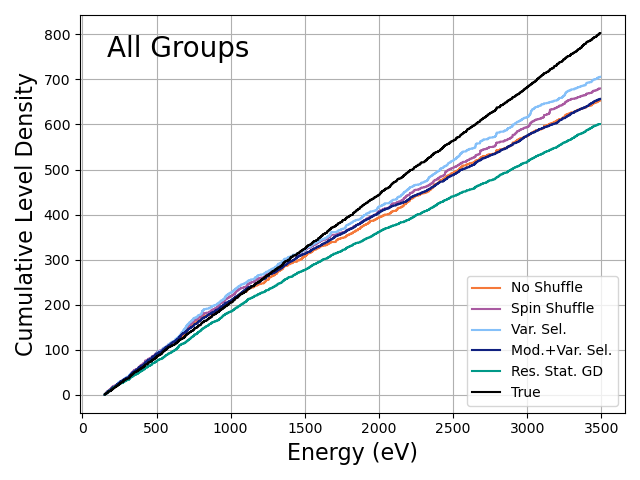}
    \caption{Empirical cumulative level density for the 3+ spin group (top left), 4+ spin group (top right), and combined spin groups (bottom) for synthetic data. Curves correspond to progressively more advanced fitting strategies, ranging from the baseline without resonance statistics to full resonance statistics -informed optimization. Spin group shuffling substantially reduces spin-assignment bias and improve agreement with the true level density.}
    \label{fig:ecdf_fake}
\end{figure}

Figure \ref{fig:ecdf_fake} shows the empirical level density for each spin group and for the combined distribution. The model without resonance statistics -informed methods (orange) is shown as a baseline. Subsequent curves represent, in order, spin group shuffling and increasingly comprehensive resonance statistics -informed fitting approaches: variable selection, model selection, and gradient-based optimization.
The baseline model without spin group shuffling exhibits a pronounced preference for the 4+ spin group at high energies -- a behavior described by a algorithmic change in Appendix \ref{ap:IFB}.
Introducing spin group shuffling substantially reduces this assignment bias by encouraging spin group populations that better reflect their expected statistical distributions. A small residual bias toward the 4+ spin group remains, primarily because the initial $\chi^2$-optimized solution is often locally optimal and preferred over shuffled alternatives. Nonetheless, spin group shuffling significantly reduces spin assignment bias relative to the models without spin group shuffling.

The fitting performance for the baseline, spin group shuffling, and resonance statistics -informed fitting models are shown in Figure \ref{fig:performance_fake_methods}. Performance metrics were evaluated over 20 eV windows across each energy region, with a 50 eV external resonance buffer on either side to account for outside resonance influence. Violin plots summarize the distributions of $\chi^2$ per data point and RMSE across all windows. The error bars on the violin plots represent the mean and standard deviation of the mean. The $\chi^2$ per data point metric for the true fit is shown in red.
At low energies, all models exhibit comparable $\chi^2$ and RMSE performance, indicating that resonance statistics -informed fitting neither improves nor degrades local fit quality.
Across all energy ranges, RMSE is largely insensitive to the introduction of spin group shuffling and resonance statistics -informed fitting. This insensitivity implies RMSE primarily measures local cross section agreement and is less sensitive to spin group assignment.

\begin{figure}
    \centering
    \includegraphics[width=0.9\linewidth]{figs/fake_performance/performance_over_energy_all_disc.png}
    \caption{Fit performance of each model applied to synthetic data. The top two rows show representative 20 eV windows across the indicated energy ranges. The third row shows violin plots of the normalized $\chi^2$ per data point, and the fourth row shows corresponding RMSE distributions, computed over 20 eV windows within each energy range. The red line indicates the $\chi^2$ obtained using the true resonance parameters.}
    \label{fig:performance_fake_methods}
\end{figure}

\begin{figure}
    \centering
    \includegraphics[width=0.45\linewidth]{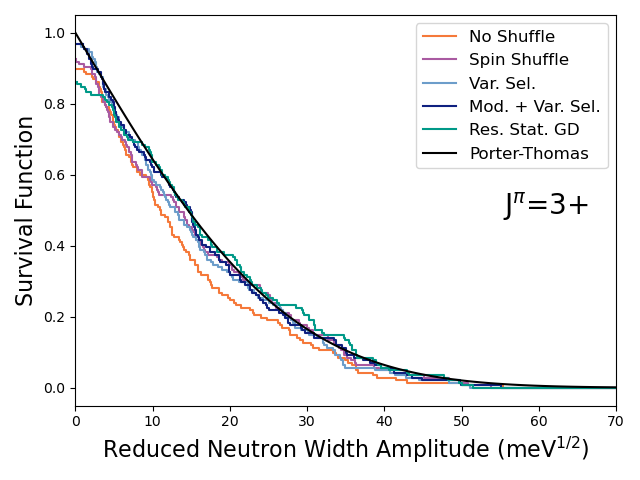}\includegraphics[width=0.45\linewidth]{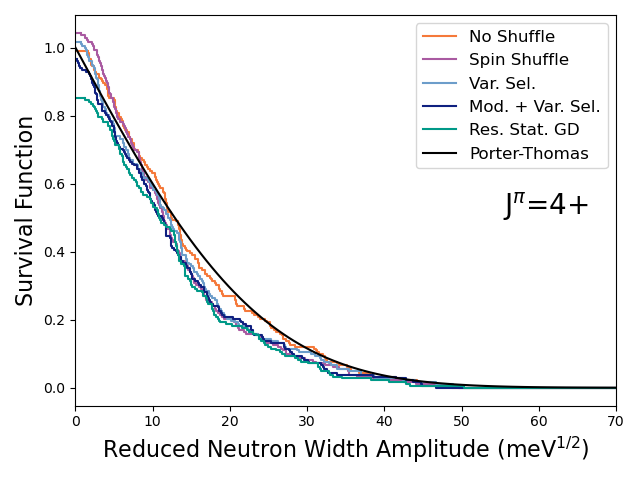}
    \caption{The Porter-Thomas empirical survival function for resonances between 150 eV and 1.5 keV for each model for the 3+ spin group (left) and the 4+ spin group (right), applied to synthetic data.}
    \label{fig:PT_ecdf_fake_1.5keV}
\end{figure}

\begin{figure}
    \centering
    \includegraphics[width=0.45\linewidth]{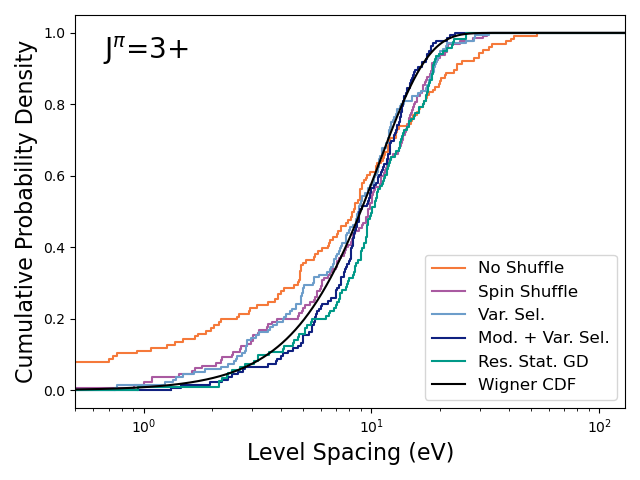}\includegraphics[width=0.45\linewidth]{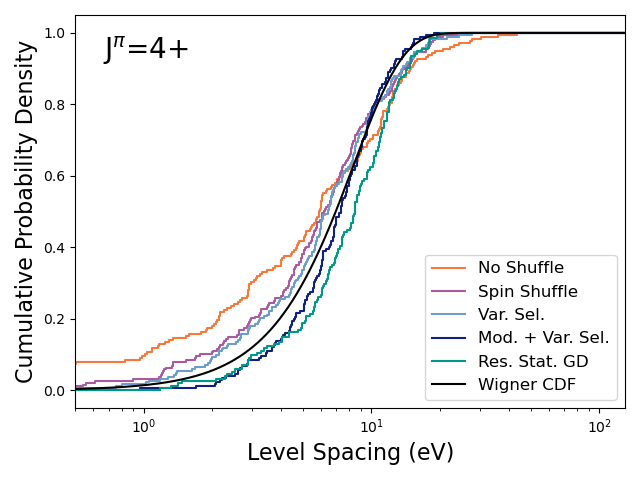}
    \caption{The empirical cumulative Wigner level spacing function for resonances between 150 eV and 1.5 keV for each model for the 3+ spin group (left) and the 4+ spin group (right), applied to synthetic data. The level spacing is plotted on a log-scale. Models with resonance statistics -informed variable and model selection show a significantly lower frequency of small level-spacings over other models.}
    \label{fig:Wig_ecdf_fake_1.5keV}
\end{figure}

Figures \ref{fig:PT_ecdf_fake_1.5keV} and \ref{fig:Wig_ecdf_fake_1.5keV} show the empirical fits to the PT and Wigner distributions for identified 3+ and 4+ resonances between 150 eV and 1.5 keV. Goodness-of-fit metrics for the PT distribution are reported in Table \ref{tab:PT_fit_fake}. 
A marginal change in PT agreement is observed across models. In contrast, the Wigner level-spacing distribution shows clear improvement when resonance statistics -informed variable and model selection are applied. Table \ref{tab:Wigner_fit_fake} quantifies this improvement for both spin groups, indicating a reduced occurrence of closely spaced same-spin resonances and improved overall agreement with expected level-spacing statistics. The performance of the resonance statistics -informed gradient descent model is primarily hindered by systematic underfitting.



\begin{table}
    \centering
    \begin{tabular}{c c | c c c c c }
         J$^\pi$ & metric & no & spin & var. & mod.+var. & res. stat. \\ 
          &  &shuffle & shuffle & sel. & sel. & GD \\ 
        \hline
    
        3+ & KS  & \textbf{0.136} & \textbf{0.098} & \textbf{0.071} & \textbf{0.048} & \textbf{0.140} \\
        3+ & CvM & 0.897 & \textbf{0.294} & \textbf{0.164} & \textbf{0.058} & \textbf{0.223} \\
        \hline
        4+ & KS  & \textbf{0.058} & \textbf{0.116} & \textbf{0.101} & \textbf{0.109} & 0.149 \\
        4+ & CvM & \textbf{0.152} & 0.856 & \textbf{0.421} & \textbf{0.633} & 1.028 \\
    \end{tabular}
    \caption{Kolmogorov-Smirnov and Cram\'er-von Mises criterions for the width distribution for each model and each spin group of the synthetic data. Metrics passing the 1\% significance threshold are bolded.}
    \label{tab:PT_fit_fake}
\end{table}





\begin{table}
    \centering
    
    \begin{tabular}{c c | c c c c c }
         J$^\pi$ & metric & no & spin & var. & mod.+var. & res. stat. \\ 
          &  &shuffle & shuffle & sel. & sel. & GD \\ 
        \hline
    
        3+ & KS  & 0.163 & \textbf{0.109} & \textbf{0.098} & \textbf{0.085} & \textbf{0.141} \\
        3+ & CvM & 1.138 & \textbf{0.407} & \textbf{0.234} & \textbf{0.193} & 0.849 \\
        3+ & AD  & 4.769 & \textbf{1.875} & \textbf{1.576} & \textbf{1.031} & 2.249 \\
        \hline
        4+ & KS  & 0.186 & \textbf{0.093} & \textbf{0.081} & \textbf{0.083} & 0.157 \\
        4+ & CvM & 1.858 & \textbf{0.669} & \textbf{0.300} & \textbf{0.278} & 1.628 \\
        4+ & AD  & 4.858 & 2.178 & \textbf{1.490} & \textbf{1.319} & 2.916 \\
    \end{tabular}
    \caption{Kolmogorov-Smirnov, Cram\'er-von Mises, and Anderson-Darling criterions for the level spacing distribution for each model and each spin group of the synthetic data. The model with resonance statistics -informed variable and model selection shows a significantly better fitting performance over other models by all metrics. Metrics passing the 1\% significance threshold are bolded.}
    \label{tab:Wigner_fit_fake}
\end{table}

\subsection{Applying Models to Real Data}

The fitting methodologies were also applied to real experimental data between 200 eV and 2.5 keV to compare side-by-side with ENDF/B-VIII.1 and the analysis presented in N. Walton's thesis \cite{Noah-thesis}. For a fair comparison, ENDF/B-VIII.1 was re-optimized locally to the same data by J. Brown using SAMMY. Many algorithmic changes have been made since Walton's analysis, the details of which are outlined in Appendix \ref{appendix}.

Applying these methods to real data will assess fitting behavior beyond idealized synthetic conditions. In contrast to synthetic data, real measurements often contain unrecognized sources of uncertainty \cite{USU-2020} and inaccurate mean resonance parameters which may negatively impact the fitting performance.

\begin{figure}
    \centering
    \includegraphics[width=0.45\linewidth]{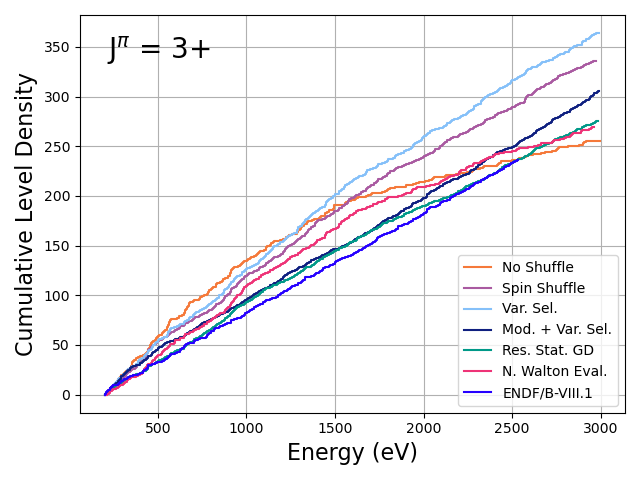}\includegraphics[width=0.45\linewidth]{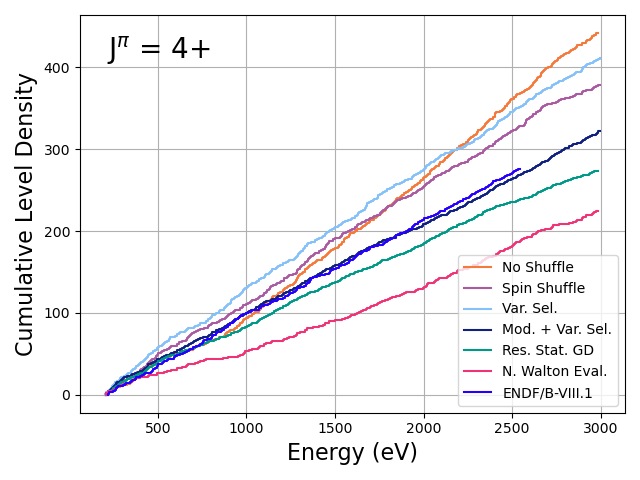}
    \includegraphics[width=0.9\linewidth]{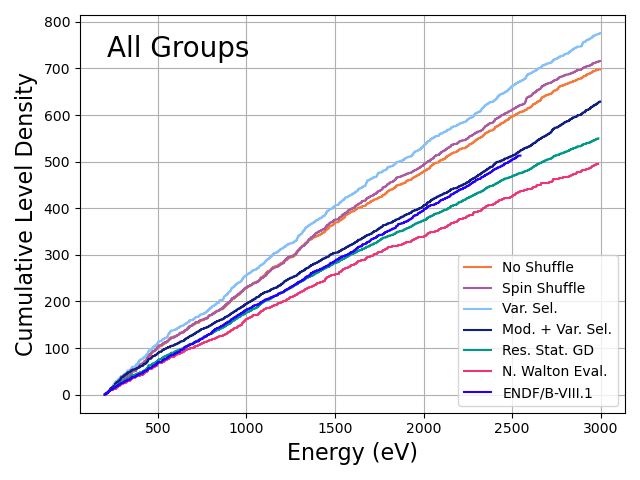}
    \caption{The empirical cumulative distribution function (ECDF) of resonance energies for real experimental data, shown for individual spin groups and their combination. Results are shown for each fitting model studied for synthetic data, along with the N. Walton fit and the ENDF/B-VIII.1 evaluation refit to the same data. Resonance statistics -informed methods suppress excessive same-spin level density relative to purely data-driven fitting.}
    \label{fig:ecdf_real}
\end{figure}

\begin{figure}
    \centering
    \includegraphics[width=0.9\linewidth]{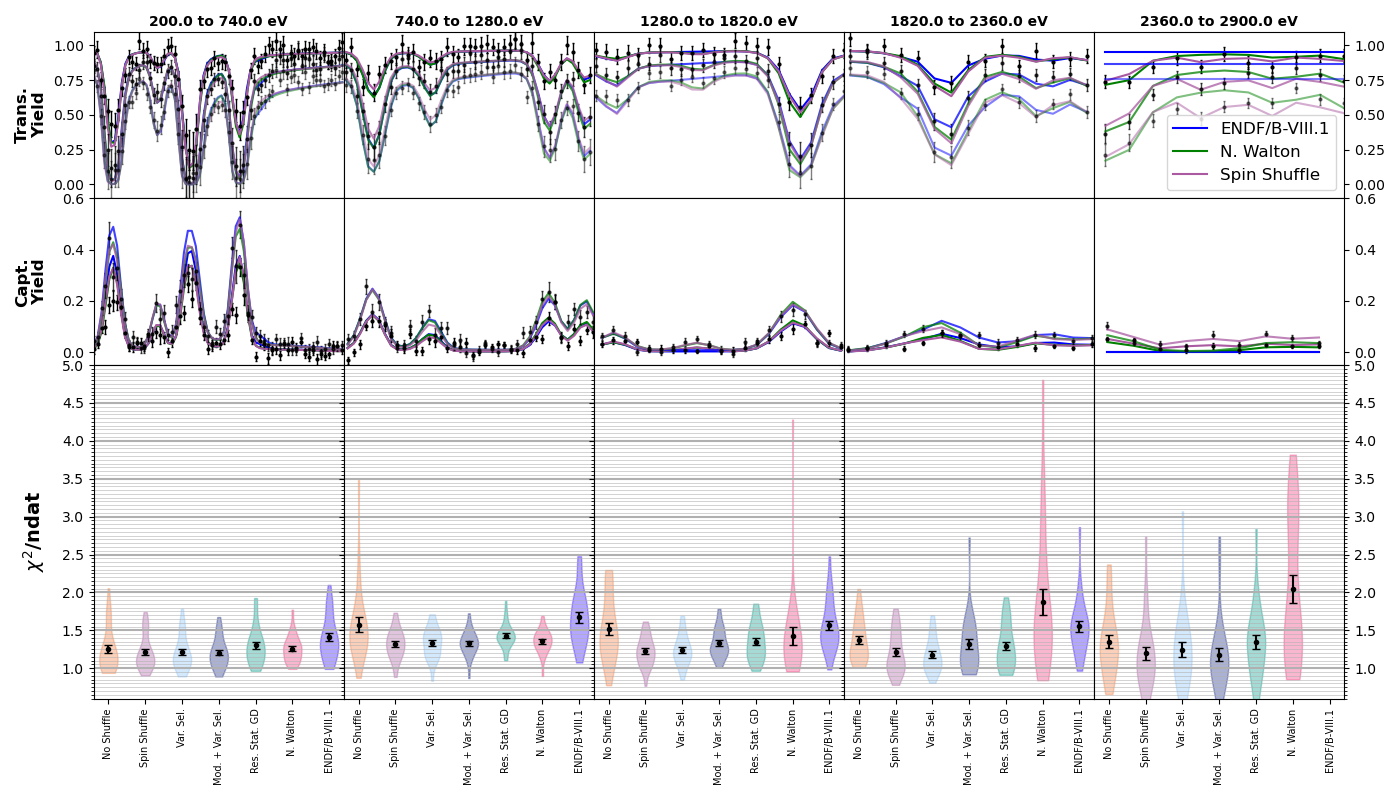}
    \caption{
    $\chi^2$ per data point for each fitting method applied to real experimental data, including a previous fit by N. Walton and ENDF/B-VIII.1 refit to the same data. Elevated $\chi^2$ values across all methods suggest unrecognized sources of uncertainty or other model deficiencies.}
    \label{fig:performance_real_methods}
\end{figure}

\begin{figure}
    \centering
    \includegraphics[width=0.45\linewidth]{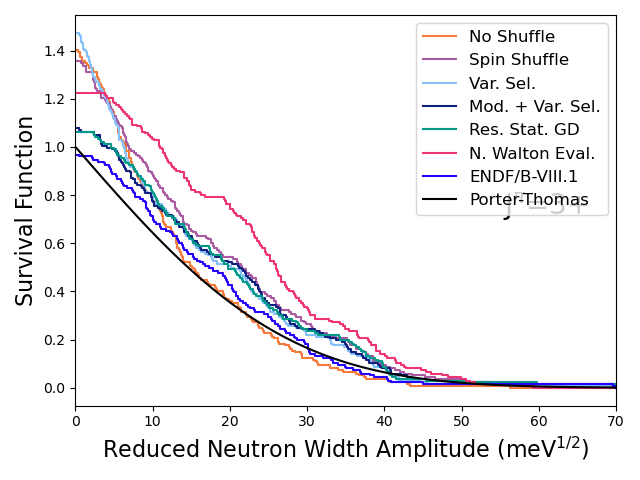}\includegraphics[width=0.45\linewidth]{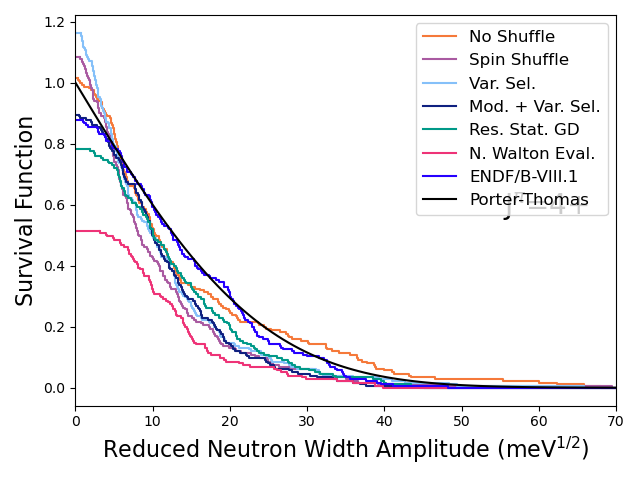}
    \caption{The empirical cumulative distribution functions of reduced neutron widths compared to the PT distribution for real experimental data between 200 eV and 1.5 keV. Deviations at small widths indicate an excess of weak resonances, consistent with overfitting. ENDF/B-VIII.1 shows the closest overall agreement with PT distribution.}
    \label{fig:PT_ecdf_real_1.5keV}
\end{figure}

\begin{figure}
    \centering
    \includegraphics[width=0.45\linewidth]{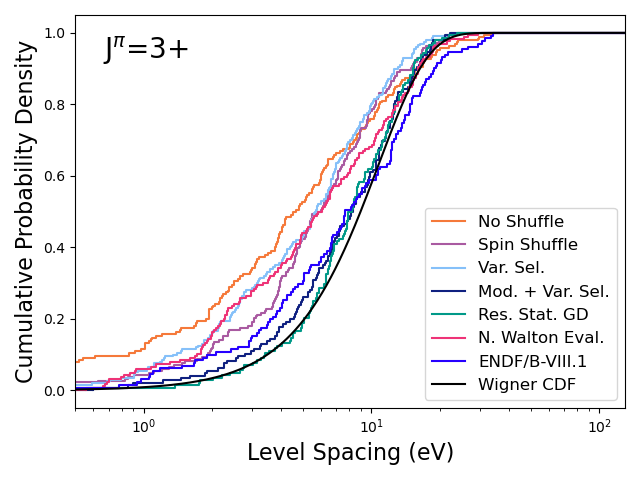}\includegraphics[width=0.45\linewidth]{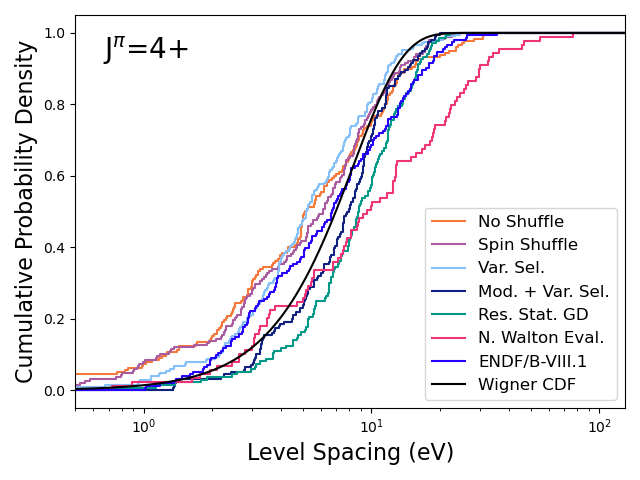}
    \caption{Empirical cumulative distribution functions of same-spin resonance spacings compared to the Wigner distribution for real experimental data between 200 eV and 1.5 keV. Resonance statistics -informed variable and model selection significantly improve agreement with Wigner spacing statistics by suppressing closely spaced same-spin resonances.}
    \label{fig:Wig_ecdf_real_1.5keV}
\end{figure}

Figure \ref{fig:ecdf_real} shows the empirical level density for each spin group and the combined distribution. Relative to ENDF/B-VIII.1, the models from this work consistently overestimate the number of resonances. This overestimation is associated with an excess of small fitted resonances, which coincides with deviations from the PT distribution for $\gamma_n<10$ meV$^{1/2}$, as shown in Figure \ref{fig:PT_ecdf_real_1.5keV}. Such behavior suggests that overfitting is driven by residual structure in the data that is not fully captured by the experimental model.
The $\chi^2$ goodness-of-fit for each model is shown in Figure \ref{fig:performance_real_methods}. Across all methods, $\chi^2$ per data point is noticeably larger for real data than for synthetic data, consistent with unreported uncertainties. Unmodeled uncertainties reduce the reliability of purely data-driven model selection and may contribute to the overfitting bias observed in the empirical level distribution in Figure \ref{fig:ecdf_real}.
Resonance statistics -informed variable and model selection partially mitigate overfitting by introducing level-spacing constraints through the Wigner distribution. As shown in Figure \ref{fig:Wig_ecdf_real_1.5keV}, these methods suppress the occurrence of closely spaced same-spin resonances, thereby reducing excessive resonance density without relying on additional experimental assumptions.

Figures \ref{fig:PT_ecdf_real_1.5keV} and \ref{fig:Wig_ecdf_real_1.5keV}, together with Tables \ref{tab:PT_fit_real} and \ref{tab:Wigner_fit_real}, quantify the agreement with PT and Wigner distributions. As observed for synthetic data, the fit to the PT distribution is largely insensitive to spin group shuffling and resonance statistics -informed fitting. ENDF/B-VIII.1 exhibits the best overall agreement with PT statistics by all three metrics.
In contrast, resonance statistics -informed variable and model selection produce the best agreement with Wigner level-spacing statistics for both spin groups for all three metrics.



\begin{table}
    \centering
    
    \begin{tabular}{c c | c c c c c c c }
         J$^\pi$ & metric & no & spin & var. & mod.+var. & res. stat. & N. Walton & ENDF \\ 
          &  & shuffle & shuffle & sel. & sel. & GD &  &  \\ 
        \hline
    
        3+ & KS  & 0.138 & \textbf{0.084} & \textbf{0.119} & 0.150 & \textbf{0.095} & 0.201 & \textbf{0.083} \\
        3+ & CvM & 0.797 & \textbf{0.179} & \textbf{0.210} & \textbf{0.507} & \textbf{0.226} & 1.484 & \textbf{0.105} \\
        \hline
        4+ & KS  & 0.135 & 0.206 & 0.214 & 0.224 & 0.142 & 0.258 & \textbf{0.075} \\
        4+ & CvM & \textbf{0.404} & 1.077 & 1.439 & 1.646 & \textbf{0.588} & 1.120 & \textbf{0.100} \\
    \end{tabular}
    \caption{Kolmogorov-Smirnov and Cram\'er-von Mises criterions for PT distribution applied to reduced neutron widths from real experimental data below 1.5 keV. Lower values indicate better agreement. The performance of the 5 models are inflated mostly by overfitting bias. PT agreement is largely insensitive to resonance statistics -informed fitting and spin group reassignment. ENDF/B-VIII.1 provides the best performance by all metrics. Metrics passing the 1\% significance threshold are bolded.}
    \label{tab:PT_fit_real}
\end{table}



\begin{table}
    \centering
    \begin{tabular}{c c | c c c c c c c }
         J$^\pi$ & metric & no & spin & var. & mod.+var. & res. stat. & N. Walton & ENDF \\ 
          &  &shuffle & shuffle & sel. & sel. & GD &  &  \\ 
        \hline
    
        3+ & KS  & 0.344 & 0.264 & 0.291 & \textbf{0.089} & \textbf{0.096} & 0.251 & \textbf{0.130} \\
        3+ & CvM & 9.325 & 5.692 & 8.222 & \textbf{0.306} & \textbf{0.237} & 3.957 & \textbf{0.696} \\
        3+ & AD  & 8.375 & 5.578 & 7.061 & \textbf{1.331} & \textbf{1.084} & 5.208 & 2.528 \\
        \hline
        4+ & KS  & 0.199 & 0.170 & 0.197 & \textbf{0.109} & 0.200 & 0.339 & 0.146 \\
        4+ & CvM & 2.222 & 1.470 & 2.645 & \textbf{0.590} & 2.588 & 2.813 & 0.886 \\
        4+ & AD  & 4.449 & 3.420 & 3.699 & \textbf{1.831} & 3.940 & inf & 3.200 \\
    \end{tabular}
    \caption{
    Kolmogorov-Smirnov, Cram\'er-von Mises, and Anderson-Darling criterions for the Wigner level-spacing distribution applied to same-spin resonances from real experimental data below 1.5 keV. Resonance statistics -informed variable and model selection yield the strongest agreement with Wigner statistics. Metrics passing the 1\% significance threshold are bolded.}
    \label{tab:Wigner_fit_real}
\end{table}

\section{Conclusion}\label{sec:conclusion}
This work studies the use of resonance statistics in cross section evaluation: spin group shuffling to address convergence issues regarding spin group assignments, and a resonance statistics -informed objective function to improve fitting performance. These methods were applied to an automated resonance fitting algorithm, where the performance was measured using established metrics.

Spin group shuffling and resonance statistics -informed fitting have small impacts on the RMSE and $\chi^2$ fitting metrics, but spin group shuffling significantly reduces spin group assignment bias above 1.5 keV.
Resonance statistics -informed variable and model selection guide resonance ladders toward consistency with Wigner spacing statistics and stabilizes the number of identified resonances when fitting to data affected by model deficiencies such as unrecognized sources of uncertainty.

Spin group shuffling with resonance statistics -informed variable and model selection is recommended for most automated evaluations. These methods improve spin group balance, reduce same-spin clustering, and produce resonance counts that are more statistically consistent when confronted with imperfect experimental data.
The resonance statistics -informed gradient descent model is unnecessary as it yields no advantage over the variable and model selection model. This model requires an external solver which incurs a large runtime penalty and is not suggested.

Future work will address average cross section bias as a performance metric. Unbiased average cross sections are important for accurate neutron transport calculations. Although originally considered for this work, this topic will be better addressed in an upcoming publication.

\section{Acknowledgments}
This material is based upon work supported by the Department of Energy National Nuclear Security Administration through the Nuclear Science and Security Consortium under Award Number(s) DE-NA0003996.

\newpage
\printbibliography

\newpage
\appendix
\section{Changes to the Automated Fitting Routine}\label{appendix}
This section discusses secondary changes to the automated fitting routine described in Walton's thesis \cite{Noah-thesis}. Many of these changes, such as the modified one-standard-error rule and stage 3 window fitting are important additions to the fitting routine, supporting the study of resonance statistics, but not critical to the hypothesis, so discussion of such features is left here for brevity.

\subsection{Modifications to the Initial Feature Bank}\label{ap:IFB}
In the initial solve phase, a dense feature bank of resonances is optimized to the objective function. The initial feature bank is seeded with many small, uniformly spaced resonances of each spin group assignment. The initial feature bank is optimized to fit the data. As optimization proceeds, resonances corresponding to physical features grow in strength as the structure of the experimental data is captured.
In the original implementation by N. Walton, initial resonances were assigned neutron widths corresponding to the 1\% quantile of the PT distribution. For $^{181}$Ta, this initialization favored growth of the 3+ spin group due to its larger neutron strength. To mitigate this bias, the present work initializes all spin groups with identical neutron widths. This modification, however, introduces bias favoring the 4+ spin group due to its larger spin statistical factor, $g_J$. Future work may mitigate spin group bias in early fitting.


\subsection{Model Selection and the Modified One-Standard-Error Rule}\label{ap:model_selection}
The model selection phase determines the final number of resonances using five-fold cross-validation (CV) \cite{Statistical-Methods}. In CV, the experimental data are partitioned into statistically independent training and validation sets. The model is fit to the training data and the objective value is evaluated on the validation data and recorded as the CV score. For each \textit{model cardinality} (i.e., number of resonances), 5 CV scores are obtained, from which a mean and standard deviation are calculated.

\begin{figure}
    \centering
    \includegraphics[width=0.5\linewidth]{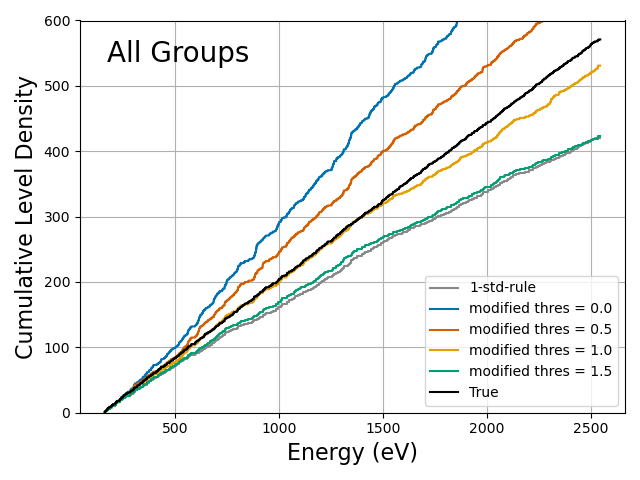}
    \caption{The cumulative level density plot for various thresholds for the modified one-standard-error rule compared against the old one-standard-error rule in gray and the true cumulative level density in black.}
    \label{fig:discrepancy_thres}
\end{figure}

Selecting the model with the optimal mean CV score tends toward overfitting due to statistical fluctuation. This bias can be rather strong at times. To reduce overfitting, N. Walton employed the \textit{one-standard-error rule}, first proposed by Breiman \textit{et al.} \cite{1-standard-error-original}. The one-standard-error rule chooses the most parsimonious model within one standard error of the minimum CV score. While effective at reducing overfitting, this rule introduces a growing bias toward underfitting as the number of data points increases -- a behavior noted by N. Walton in appendix A.3 of his thesis \cite{Noah-thesis}. 
This work adopts the \textit{modified one-standard-error rule}, described by L. Yates \textit{et al.} \cite{1-standard-error-correction}, which resolves the underfitting. The models for each cardinality share the same data and are therefore correlated, but the original one-standard-error rule neglects this correlation, motivating the modified rule. The modified one-standard-error rule is analogous to the original one-standard-error rule: select the most parsimonious model that is not \textit{discrepant} with the optimal model.
The correlation-corrected discrepancy $\Delta_{ij}$ between models $i$ and $j$ is defined in equations \eqref{eq:mean_for_discrepancy}, \eqref{eq:var_for_discrepancy}, and \eqref{eq:discrepancy}. $d_{ij,k}$ is defined as the difference between CV scores of models $i$ and $j$, for train/validation set $k$.
\begin{equation}\label{eq:mean_for_discrepancy}
\hat{d}_{ij} = \frac{1}{K}\sum_{k=1}^K d_{ij,k}
\end{equation}
\begin{equation}\label{eq:var_for_discrepancy}
\left(\delta\hat{d}_{ij}\right)^2 = \frac{1}{K(K-1)}\sum_{k=1}^K\left(d_{ij,k}-\hat{d}_{ij}\right)^2
\end{equation}
\begin{equation}\label{eq:discrepancy}
\Delta_{ij} = \frac{\left|\hat{d}_{ij}\right|}{\delta\hat{d}_{ij}}
\end{equation}
The choice threshold of $\Delta_{ij}$ to be considered discrepant is arbitrary. Larger thresholds favor more parsimonious models at the risk of missing small resonances, while smaller thresholds recover more weak resonances at the cost of increased false resonances.
The resonance statistics distributions depend critically on an unbiased estimator of the number of resonances. Consequently, a discrepancy threshold that provides an approximately unbiased estimate of the level density was selected for this work.
Figure \ref{fig:discrepancy_thres} shows the empirical cumulative level density obtained using several discrepancy thresholds. The original one-standard-error rule significantly underestimates the number of resonances. 
The modified one-standard-error rule with a threshold of 1.0 provides an excellent estimate of the true level density, particularly for energies below 1.5 keV. While there is nothing fundamentally special about a threshold value of 1.0, it performs well for the present data.

\subsection{Windowed Resonance Evaluation and Window Stitching}\label{sec:windows}
\begin{figure}
    \centering
    \includegraphics[width=0.5\linewidth]{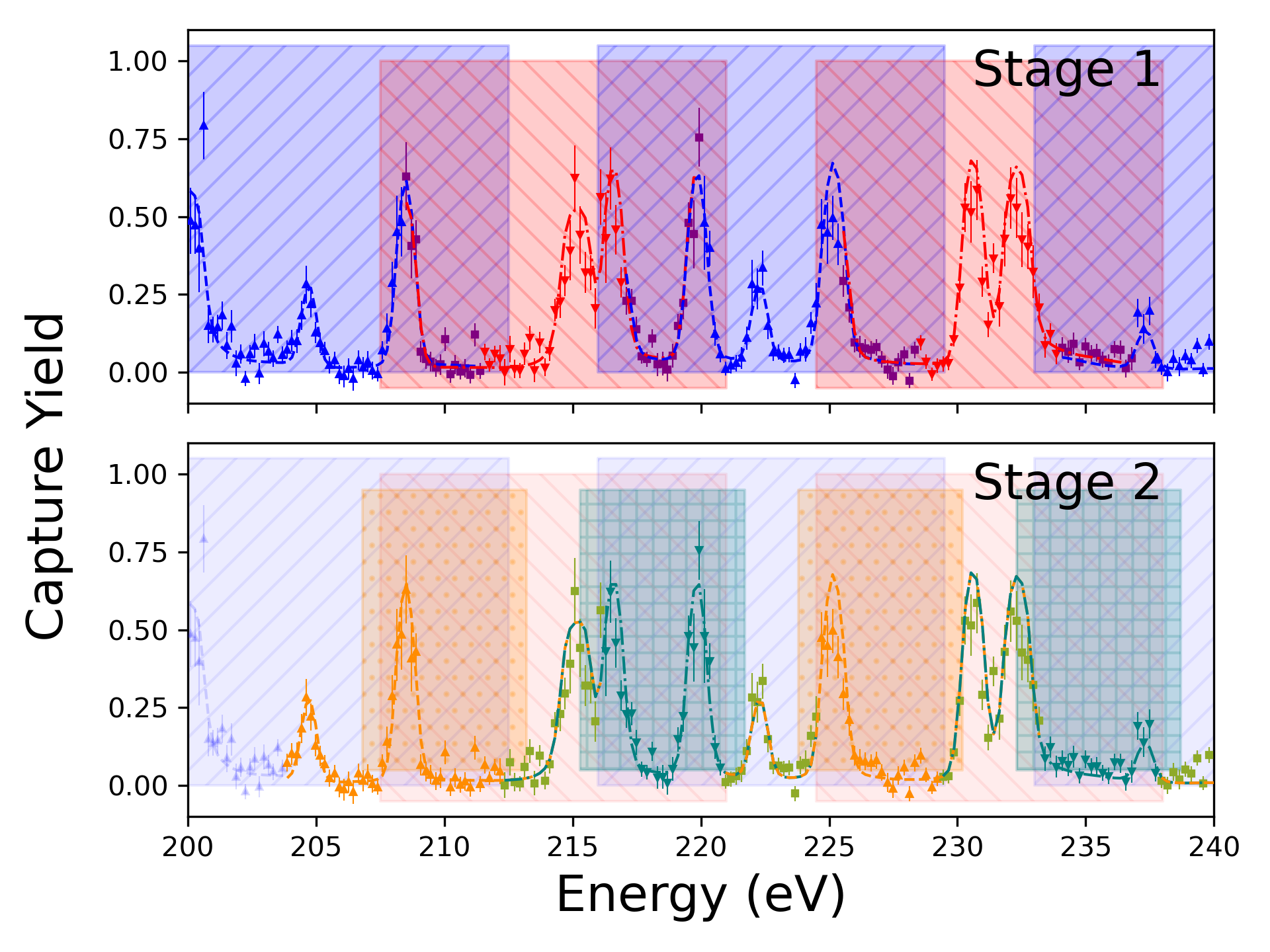}
    \caption{A diagram illustrating stage 1 (top) and stage 2 (bottom) of the window stitching algorithm. The shaded regions show the region where resonances are allowed to vary. The color and marker style show extent of the data used in the window. Regions where the data overlap is given by purple square markers in stage 1 and green square markers in stage 2.}
    \label{fig:stages_1_and_2}
\end{figure}

The initial feature bank optimization is computationally intensive due to the large number of candidate resonances. As a result, the full fitting framework is applied only to small energy windows, typically containing no more than ten resonances. Realistic evaluations may involve thousands of resonances across the RRR. To address the computational limitation, a windowed evaluation strategy is employed, as described in Walton’s thesis \cite{Noah-thesis}.
The windowed approach has two main stages, illustrated in Figure \ref{fig:stages_1_and_2}. In stage 1, resonances are fit on many small overlapping windows spanning the RRR. In stage 2, overlapping solutions are combined, and redundant resonance features are eliminated through an additional elimination process. The resulting resonance parameters are merged to construct a continuous resonance ladder over the full energy range.

This work uses 11.5 eV windows for stage 1 that have a 3 eV overlap with each neighboring window. There is a 1 eV buffer where resonances are fit, but data are not used.
Stage 2 windows, centered on each overlapping region, are 6.4 eV in size. Outside the stage 2 window, fixed resonances are taken from neighboring stage 1 windows; these resonances are not optimized, nor included in the final model, but they capture background resonance contributions. There is also a 3 eV buffer where additional data is included so resonances near the edge of the window can optimize with more data.

\begin{figure}
    \centering
    \includegraphics[width=0.65\linewidth]{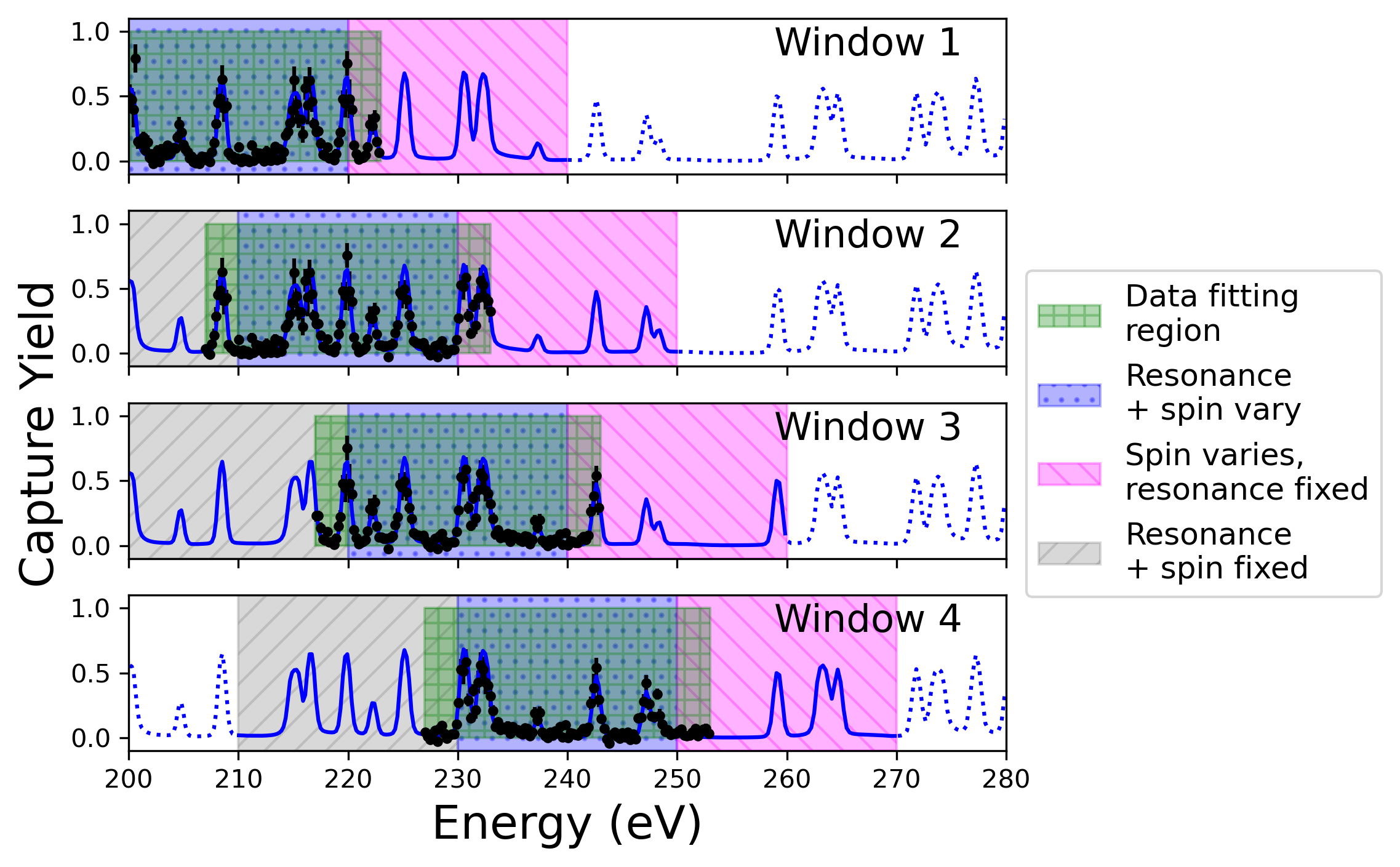}
    \caption{A schematic of the stage 3 evaluation. The data in the green gridded region are fit using resonance parameters in the other shaded regions. Spin group assignments are shuffled and resonances are optimized in each window sequentially. Windows are reevaluated from left to right until the last window has been reevaluated. In the blue-shaded region (center), resonance parameters and spin group assignments are allowed to vary. In the magenta-shaded region (right), resonance parameters are fixed, but spin group assignments are allowed to vary. In the gray-shaded region (left), both resonance parameters and spin group assignments are fixed.}
    \label{fig:stage_3}
\end{figure}

At low energies, resonance features are localized and well-suited to small-window fitting. However, level-spacing statistics span larger energy intervals, and at higher energies resonance widths often approach the scale of the fitting windows. In these regimes, resonances are optimized using only a subset of the full dataset.
To address these limitations, a third fitting stage is introduced. In this stage, the resonance ladder is reevaluated using larger windows, with spin group assignments reshuffled and resonance parameters refit in an iterative, low-to-high energy sequence.
Figure \ref{fig:stage_3} illustrates the stage 3 fitting procedure. The window sizes in the figure are consistent with implementation. This refitting stage is applied to all models considered in this work.

\subsection{Fitting Hyperparameters and Resonance Properties}
For consistency, this work uses optimizer settings outlined in table B.3 of N. Walton's thesis \cite{Noah-thesis}; the exception is the expansion of the 11 eV window to an 11.5 eV window, described in Section \ref{sec:windows}.
Mean resonance parameters used for resonance statistics are provided in table B.2 of N. Walton's thesis.

\end{document}

%% file: formatting.tex
\usepackage[T1]{fontenc}
\usepackage[utf8]{inputenc}
\usepackage[english]{babel}

\usepackage{amsmath}
\usepackage{amsfonts}
\usepackage{amsthm}
\usepackage{mathtools}

\usepackage{graphicx}
\graphicspath{{./images/}}
\usepackage{booktabs}
\usepackage{caption}
\usepackage{wrapfig}

\usepackage{algorithm}
\usepackage{algpseudocode}

\usepackage[dvipsnames]{xcolor}
\definecolor{revise}{HTML}{990000}

\usepackage{microtype}

\usepackage{noto}

\usepackage[
  left=30mm,
  right=30mm,
  top=25mm,
  bottom=25mm
]{geometry}

\usepackage[
  backend=biber,
  style=numeric,
  sorting=none
]{biblatex}
\usepackage{hyperref}
\hypersetup{
  colorlinks=true,
  linkcolor=black,
  citecolor=black,
  urlcolor=black
}

\renewenvironment{abstract}
 {\small\begin{center}\bfseries\abstractname\end{center}\quotation}
 {\endquotation}
